\documentclass[useAMS]{mn2e}
\usepackage{amsmath}
\usepackage{times}
\usepackage{graphicx}

\title[Light curve evolution of Swift J1357.2-0933]
{Multi-wavelength light curve evolution of Swift J1357.2-0933 during its 2011
outburst}

\author[Weng \& Zhang]{Shan-Shan Weng$^{1,2,3}$\thanks{E-mail:
wengss@ihep.ac.cn} \thanks{Current address: Department of Physics and Institute
of Theoretical Physics, Nanjing Normal University, Nanjing 210023,
China} Shuang-Nan Zhang$^{2,4,5}$\thanks{E-mail: zhangsn@ihep.ac.cn}\\
$^1$\,Sabanc\i~University, Faculty of Engineering and Natural
Sciences, Orhanl\i$-$ Tuzla, \.{I}stanbul 34956, Turkey \\
$^2$\,Key Laboratory of Particle Astrophysics, Institute of High Energy
Physics, Chinese Academy of Sciences, Beijing 100049, China \\
$^3$\,Department of Physics, Xiangtan University, Xiangtan 411105, China \\
$^4$\,National Astronomical Observatories, Chinese Academy of Sciences,
Beijing, 100012, China \\
$^5$\,Physics Department, University of Alabama in Huntsville, Huntsville, AL 35899, USA \\
}

\date{}

\pagerange{\pageref{firstpage}--\pageref{lastpage}} \pubyear{2014}

\begin{document} \maketitle \label{firstpage}

\begin{abstract}
Swift J1357.2-0933 underwent an episodic accretion in 2011 and provided very
regular temporal and spectral evolution, making it an ideal source for
exploring the nature of very faint X-ray transients (VFXTs). In this work, we
present a detailed analysis on both X-ray and near-ultraviolet (NUV) light
curves. The fluxes at all wavelengths display a near-exponential decays in the
early phase and transits to a faster-decay at late times. The e-folding decay
time-scales monotonically decrease with photon energies, and the derived
viscous time-scale is $\tau_{\rm \dot{M}} \sim 60$ days. The time-scale in the
late faster-decay stage is about a few days. The high ratio of NUV luminosity
to X-ray luminosity indicates that the irradiation is unimportant in this
outburst, while the near-exponential decay profile and the long decay
time-scales conflict with the disc thermal-viscous instability model. We thus
suggest that the disc is thermally stable during the observations. Adopting the
truncated disc model, we obtain a lower limit of peak accretion rate of $0.03
\dot{M}_{\rm Edd}$ and the X-ray radiative efficiency $\eta < 5\times10^{-4}$,
which decreases as the luminosity declines. The low X-ray radiative efficiency
is caused by the combined action of advection and outflows, and naturally
explains that the X-ray reprocessing is overwhelmed by the viscous radiation of
the outer standard disc in the NUV regime. We also propose a possibility that
the outer standard disc recedes from the central black hole, resulting in the
faster-decay at late times.
\end{abstract}

\begin{keywords}
accretion, accretion discs --- black hole physics --- X-rays: binaries ---
X-rays: stars --- X-rays: individual (Swift J1357.2-0933)
\end{keywords}

\section{Introduction}

Low-mass X-ray binaries (LMXBs) are systems in each of them a low-mass
companion transfers material via Roche-lobe overflow onto a black hole (BH) or
a weakly magnetized neutron star (NS). They spend most time in a dim, quiescent
state with a X-ray luminosity of $10^{31-34}$ erg s$^{-1}$, but occasionally
they undergo bright X-ray and optical outbursts, which are ascribed to a huge
increase of accretion rates. Based on their peak X-ray luminosity ($L_{\rm
X}^{\rm peak}$ in 2--10 keV), LMXBs can be classified into three subclasses:
bright ($L_{\rm X}^{\rm peak} \sim 10^{37-39}$ erg s$^{-1}$), faint ($L_{\rm
X}^{\rm peak} \sim 10^{36-37}$ erg s$^{-1}$), and very faint X-ray transients
($L_{\rm X}^{\rm peak} \sim 10^{34-36}$ erg s$^{-1}$; Wijnand et al. 2006). At
present, it is widely accepted that outbursts of bright X-ray transients arise
from the disc thermal-viscous instability (Chen et al. 1997), which also offers
promising explanation for other accretion systems, e.g., dwarf novae and active
galactic nuclei. However, the standard disc instability model (DIM) is unable
to reproduce the observed long recurrence times nor the typical ``fast-rise
exponential-decay'' light curves; moreover, it produces multiple re-flares
which have never been observed (see Lasota 2001 for reviews). There have been
several attempts to modify the standard DIM to account for these observed
properties, often by invoking irradiation heating (King \& Ritter 1998), low
$\alpha$-viscosity (Menou et al. 2000), and truncated disc (Dubus et al. 2001).
Taking both the irradiation and truncation effects into account, Dubus et al.
(2001) argued that the outburst decay is divided into three stages. First, the
X-ray irradiation inhibits the disc from returning to the cool state; the decay
is viscous as the thermal equilibrium can be maintained. The light curve shows
a near-exponential decay. Second, the accretion rate becomes low enough that
the irradiation is not sufficient to ionize the outer edge of the accretion
disc; however, the decay rate is still controlled by the irradiation. The light
curves exhibit a linear decay. In the final stage, the irradiation plays no
role and light curves quickly decay on a thermal time-scale.

To date, dozens of very faint X-ray transients (VFXTs) have been discovered
(Muno et al. 2005). A significant fraction of them have exhibited type I X-ray
bursts (Cornelisse et al. 2002; Degenaar \& Wijnands 2009), and their
companions are fainter than B2 IV stars since no optical and infrared
counterpart was detected (Muno et al. 2005). Therefore, it is likely that these
sources are LMXBs, and the unusually low outburst luminosity might imply very
low time-averaged mass accretion rates in VFXTs. In the DIM framework, if the
mean mass transfer rates are much larger than $10^{-14} M_{\odot}$ yr$^{-1}$,
the duty cycles of transients will be larger than those observed, i.e.,
frequent outbursts are expected for most sources (Maccarone \& Patruno 2013).
However, King \& Wijnands (2006) pointed out that the mass of a companion can
only lose a mass of $< 10^{-3} M_{\odot}$ within a Hubble time if the mean mass
transfer rates are less than $10^{-13} M_{\odot}$ yr$^{-1}$. That is, the
standard LMXBs evolution is too slow for normal stellar-mass companions to
reach the very low mass stars ($< 0.1 M_{\odot}$) that probably exist in VFXTs,
posing a challenge to the binary evolution theory (King \& Wijnands 2006).

The nature of VFXTs is still a puzzle, and our study on VFXTs is hampered by
several factors. Due to their relatively faint radiations, the most sensitive
X-ray instruments are required to collect good quality spectra and determine
their outburst durations. A large number of VFXTs are found close to Sgr
A$^{\ast}$ owing to XMM-Newton/Chandra/Swift monitoring campaigns in the
Galactic centre region (e.g., Wijnands et al. 2006; Degenaar \& Wijnands 2009).
However, near-ultraviolet (NUV) \footnote{In this paper, we use the term
``NUV'' to include also the V, B, and U bands.} detection is not feasible in
the direction of the Galactic centre due to the serious absorption. The
outburst durations of VFXTs span from days to months (Degenaar \& Wijnands
2009). It is worth noting that we hardly know when outbursts are exactly over
due to instrumental limitations, and the values of their outburst durations
also depend on the instrument sensitivity. The decay time-scale is a better
parameter to describe light curves; however, it is more difficult to estimate
because the stochastic variations and irregular profiles usually presented in
the light curves of VFXTs. As a result, there is still lack of light curve
analysis for VFXTs.

The VFXT, Swift J1357.2-0933, is located at the position of $l = 328.702\degr$
and $b = +50.004\degr$ (Galactic coordinate), and was first detected by {\it
Swift} on 2011 January 28 during its outburst with a low peak luminosity $\sim
10^{35}$ erg s$^{-1}$ (Krimm et al. 2011). The short distance ($D \sim 1.5$
kpc; Rau et al. 2011) and low extinction in its direction (high Galactic
latitude) allow us to obtain a set of NUV data, which had not been available
for any VFXT before. Armas Padilla et al. (2013) investigated the correlations
between the simultaneous X-ray and NUV luminosities, and argued that its NUV
emissions originate from a viscously heated disc instead of the X-ray
reprocessing. Analyzing its spectroscopic observations, Corral-Santana et al.
(2013) reported that the orbital period of the binary is $2.8\pm0.3$ h and the
mass of the compact star exceeds 3 $M_{\odot}$, making it the first BH VFXT.
Its companion is very red, and is estimated to be an unevolved M4 star (Rau et
al. 2011). Using the empirical relation between the outburst amplitude and the
orbital period of LMXBs, Shahbaz et al. (2013) estimated the $V$-magnitude of
the companion in quiescence to lie in the range $V_{\rm min} = 22.7$ to 25.6.
Assuming that the secondary star is a M4.5 star, they gave a distance $D =
0.5-6.3$ kpc based on the distance modulus.

The X-ray and NUV light curves of Swift J1357.2-0933 during its 2011 outburst
exhibited relatively simple exponential decay profiles, which are worth a
careful study. In this paper, we firstly estimate some basic parameters (e.g.,
the decay time-scale and the accretion rate), then discuss the origin of its
NUV emission, and test the DIM. The {\it Swift} data reduction is described in
the next section. The results and their physical implications are presented in
Sections 3 and 4. Conclusion and Discussion follow in Section 5.

\section{Data Reduction}

{\it Swift} is a multi-wavelength observatory with three scientific instruments
on board: the Burst Alert Telescope ({\it BAT}), the X-ray Telescope ({\it
XRT}), and the UV/Optical Telescope ({\it UVOT}). Swift J1357.2-0933 triggered
the {\it BAT} on 2011 January 28 (Krimm et al. 2011), and then {\it Swift}
executed 43 pointings in the following 7 months. We process both the {\it XRT}
and the {\it UVOT} data with the packages and tools available in HEASOFT
version 6.14.

The {\it XRT} data were taken in photon-counting (PC) mode before February 1
and after May 13, and the window-timing (WT) mode was used between February 1
and May 13. The initial event cleaning is performed with the task {\it
xrtpipeline} using standard quality cuts. The source light curves are extracted
within a circle of radius 25 pixels centreed at the source position with {\it
xselect}, while an annulus region with the radius 25 and 50 pixels is adopted
for background region. The first observation taken in PC mode is strongly
affected by pileup, and thus excluded in this work. The light curves are
corrected with the task {\it xrtlccorr} accounting for telescope vignetting and
point spread function corrections due to the geometry of the light curve
extraction region. The scaled background rate is then subtracted from the
corrected light curves. The net count rates are averaged for each observation,
and we also use the spectral fitting results published in Armas Padilla et al.
(2013).

The {\it UVOT} has six filters: {\it V}, {\it B}, {\it U}, {\it UVW1}, {\it
UVM2} and {\it UVW2} with a coverage of 1700--6000 \AA. The {\it UVOT}
observations were performed in the image mode. When available, the sky image is
summed for each observation with {\it uvotimsum} in order to increase photon
statistics. We perform aperture photometry with the summed images by using {\it
uvotsource} with an aperture radius 5{\arcsec}, and the background flux density
is extracted from a neighboring source-free sky region. The NUV flux is
calculated by assuming the constant flux density within each filter band, which
is read from their response file (version 105). We also correct the integral
luminosity in {\it UVOT} band (2.1--7.8 eV) for the Galactic extinction
following the description in Armas Padilla et al. (2013).


\section{Light curve evolution}

As shown in Figure 1, the fluxes at all wavelengths decrease monotonically and
track each other consistently until below the instrument detection limit after
$\sim 200$ days. Figure 1 also shows that the light curves show
near-exponential decay in the early phase. We fit the light curves (either
count rates or flux) with the exponential function
\begin{equation}
flux = C*\exp(-t/\tau),
\end{equation}
where $C$ is a constant, $t$ is observational date, and $\tau$ is the decay
time-scale. The time-scales monotonically decreases with photon energies
ranging over four orders of magnitude (Figure 2). 
The e-folding decay time of X-ray is $\tau_{\rm X}\sim 30$ days, while the
time-scales of longer-wavelength light curves are $\tau_{\rm NUV}\sim 80-150$
days, i.e., $\tau_{\rm X}/\tau_{\rm NUV}\sim 0.2-0.38$.

\begin{figure*}
\includegraphics[width=18cm]{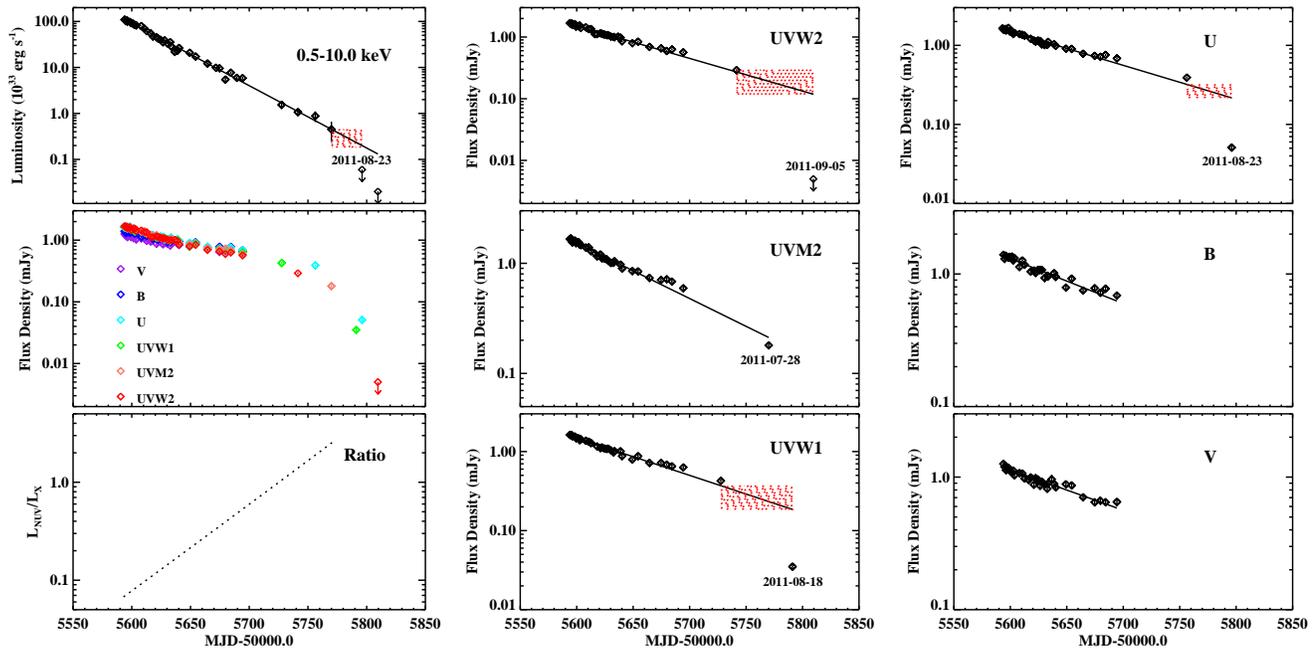}
\caption{The light curves (diamond points) at different wavelengths are fitted
with the exponential function (solid lines). The red boxes mark the time and
luminosity intervals of light curve transitions. The ratio of NUV luminosity to
X-ray luminosity ($\frac{L_{\rm NUV}}{L_{\rm X}}$, dotted line) is calculated
with the assumption of exponential decay. \label{lightcurve}}
\end{figure*}

\begin{figure}
\includegraphics[width=9cm]{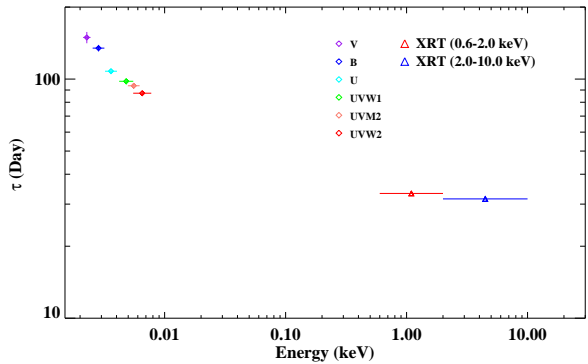}
\caption{The decay time-scale $\tau$ for different energy bands.\label{decay}}
\end{figure}

Extrapolating the solid line in Figure 1, we expect a X-ray luminosity $L_{\rm
X} \sim 1.8 \times 10^{32}$ erg s$^{-1}$ (in 0.5--10.0 keV) on 2011-08-23,
corresponding to an absorbed flux of $8.6 \times 10^{-13}$ erg cm$^{-2}$
s$^{-1}$ (in 0.2--10.0 keV), which is above the {\it XRT} sensitivity
\footnote{http://swift.gsfc.nasa.gov/about\_swift/xrt\_desc.html}. However,
{\it XRT} did not detect the source in the last two observations, indicating
that the light curves transit to a faster-decay at late times. The transition
should happen between the last observation (2011-07-28) that agrees with the
exponential fitting (the solid line in Figure 1) and the first observation
(2011-08-23) that deviates from the fitting. The transition luminosity is
deduced from the observational time of these two data assuming the exponential
decay. We suggest that the transition occurred at the X-ray luminosity of a few
times $10^{32}$ erg s$^{-1}$ between 2011-07-28 and 2011-08-23 (red boxes
region in Figure 1) assuming a distance of 1.5 kpc (Rau et al. 2011).

The {\it UVOT} data also exist the similar transition. The integral luminosity
in {\it UVOT} band (2.1--7.8 eV) decreases from the peak value of
$7.8\times10^{33}$ erg s$^{-1}$ to $2.9\times10^{33}$ erg s$^{-1}$ on
2011-05-13, after which time the observations were sparse and only one filter
was used for each observation. The {\it UVM2} observation on 2011-07-28 still
follows (at least not significantly deviates from) the exponential decay,
indicating that the transition takes place after 2011-07-28. The {\it UVW1}
observations further suggest that the transition happened between 2011-07-28
and 2011-08-18, that is, the light curves of X-ray and NUV might transit almost
simultaneously. The NUV light curves present relatively shallower decay, and
the NUV luminosity within all six filter bands at the onset of transition is
estimated to be $L_{\rm NUV} \sim (0.9-1.1)\times10^{33}$ erg s$^{-1}$
according to the exponential law. Even though entering into a faster-decay
phase after 2011-08-18, the source can still be detected by the {\it U} filter
on 2011-08-23, implying that the decay time-scale is of a few days at late
times.

Based on the exponential decay scenario, we calculate the NUV luminosity
($L_{\rm NUV}$) within 2.1-7.8 eV and the ratio of $L_{\rm NUV}$ to $L_{\rm X}$
(Figure 1). The ratio increases up to 2 to 3 as outburst declines at the onset
of light curve deviation from the exponential decay law. When entering into the
late faster-decay stage, the fluxes of the last observations with {\it UVW1}
filter (2011-08-11) and {\it U} filter (2011-08-23) are lower than those
predicted by exponential law by a factor of $\sim 4-6$, putting a lower limit
of $\frac{L_{\rm NUV}}{L_{\rm X}}\geq0.5$.

\section{Radiatively inefficient accretion flows}

The origins of NUV emissions in XRBs are diverse. The strong correlation
between the X-ray flux and the NUV flux $L_{\rm NUV} \propto L_{\rm X}^{\beta}$
has been revealed in several sources. Using the {\it Swift} monitoring data of
XTE J1817--330, Rykoff et al. (2007) showed the correlation slope $\beta = 0.47
\pm 0.02$, that is consistent with the value predicted by the X-ray
reprocessing model. In contrast, the small values of $\beta \sim 0.2-0.38$
observed in the 2011 outburst of Swift J1357.2-0933 implies that the X-ray
irradiation contributes little or no NUV emission (Armas Padilla et al. 2013).
In this work, we further completely rule out the irradiation hypothesis since
the NUV luminosity is close to and even exceeds the X-ray luminosity
($\frac{L_{\rm NUV}}{L_{\rm X}} \sim 1$) at late times. On the other hand, the
companion is very red and contributes tiny radiations in {\it UVOT} band with a
$V$-magnitude of $V_{\rm vin} = 22.7-25.6$ (Shahbaz et al. 2013). So we confirm
that the viscously heated disc is the only option for the NUV emission.

In LMXBs the inner disc is hot, once the accretion rate is below the critical
value, the outer disc becomes too cold ($< 10^{4}$ K) and somewhere in between
the disc must be unstable. The reason is that in this region the accreted
matter is partially ionized, and the hydrogen recombination results in a large
change in opacity, affecting the thermal equilibrium. If the instability
propagates in the disc, it would trigger re-flares, which are not present in
the data of Swift J1357.2-0933. It was widely believed that the irradiation
plays a dominant role in LMXB behaviors, e.g., keeping the disc ionized and
stabilizing it. Menou et al. (2000) argued that the re-flares can be prevented
without the irradiation, if the inner and most unstable part of the accretion
disc was replaced by a hot accretion flow and the viscosity parameter was
significantly small. However, their model failed to reproduce several important
features of light curves (Figure 8 in their paper): 1) The output light curves
are markedly different from the exponential shape; 2) The decay time-scales
($\sim 10$ days) are too short; 3) In the final stage, the light curves decay
steeply on a thermal time-scale. Lasota (2001) suggested that re-flares were a
fundamental property of DIM. The exponential decay profiles without re-flares
shown in the outburst of Swift J1357.2-093313 can not be comprehended in the
framework of DIM when the irradiation is negligible. In addition, the decay
time-scale of a few days in the late faster-decay stage is much longer than the
thermal time-scale.

Alternatively, we suggest that the outburst is driven by a stable viscous
process, and the whole disc is fully ionized during the {\it Swift}
observations. Assuming that the accretion disc radius $R$ and the kinematic
viscosity $\nu$ are nearly constant, the surface density is quasi-steady and
goes as $\Sigma \simeq -\frac{\dot{M}}{3\pi\nu}$ (King \& Ritter 1998; Frank et
al. 2002). The mass of the disc can be given by integrating the surface
density: $M = 2\pi\int^{R}_{0}\Sigma R dR\simeq -\dot{M}\frac{R^{2}}{3\nu}$,
that is, the accretion rate decays as $\dot{M} \propto exp(-3\nu t/R^{2})$.
Thus, the light curves would naturally draw out exponential profiles if the
luminosity scales as $L \propto \dot{M}^{b}$.

Investigating the motions of its H$\alpha$ emission line wings and the
double-peak separation, Corral-Santana et al. (2013) suggested that the orbital
period of the Swift J1357.2-093313 is $2.8$ h, a large inclination $i \sim
70\degr$, and the mass of the black hole exceeds 3 $M_{\odot}$. The short
orbital period points to a small accretion disc, $R_{\rm disc} \sim$ a few
times $10^{10}$ cm, and the unusually low $L_{\rm X}^{\rm peak} \sim 10^{35}$
erg s$^{-1}$ corresponds to $\xi \sim 10^{-4}$ Eddington luminosity of a $10
M_{\odot}$ black hole. Both theoretical and observational works suggested that
$L_{\rm X}^{\rm peak}$ increases with orbital periods (e.g., King 2000; Wu et
al. 2010). Analyzing a large sample of LMXBs, Knevitt et al. (2014) claimed
that the $L_{\rm X}^{\rm peak}$ of LMXBs with orbital periods of $< 4$ h would
be lowered owing to the transition to the radiatively inefficient accretion
flows (RIAFs). Previous X-ray spectral fitting shows an anti-correlation
between the photon index ($\Gamma$) and the X-ray luminosity (Armas Padilla et
al. 2013), which is also found in some other XRBs (Kalemci et al. 2005, 2013;
Corbel et al. 2006). Wu \& Gu (2008) interpreted this relationship as due to
the RIAFs. It means that the outer standard disc is truncated at a transition
radius ($R_{\rm tr}$) by an inner hot accretion flow (see Zhang 2013; Yuan \&
Narayan 2014 for reviews). Qiao \& Liu (2013) employed the truncated disc model
to study the relation between $\Gamma$ and $\xi$ in the sub-Eddington
accretion. They argued that at accretion rates ($\dot{M}$) lower than $\sim
0.01$ Eddington accretion rate, the inner disc vanishes completely by
evaporation, and the accretion is dominated by inner advection-dominated
accretion flows (ADAFs, Narayan \& Yi 1995), which can produce the
anti-correlation between $\Gamma$ and $\xi$. Nowadays, it is widely believed
that the outer thin disc is restricted beyond a large radius $R_{\rm tr}$, and
the radiation is extremely inefficient at low accretion rates (e.g., Esin et
al. 1997; Fender et al. 2004).

For RIAFs, the X-ray luminosity scales as $L_{\rm X} \propto \dot{M}^{2.0}$
(Russell et al. 2006; Shahbaz et al. 2013) and so $\tau_{\rm X} \sim 0.5
\tau_{\rm \dot{M}}$. If the NUV luminosity originates in the viscous thin disc
with temperature is $\sim 10^4$ K, the expected correlation in the NUV is
$L_{\rm NUV} \propto \dot{M}^{0.5}$ and so $\tau_{\rm NUV} \sim 2 \tau_{\rm
\dot{M}}$ (Frank et al. 2002). We therefore have $\tau_{\rm X}/\tau_{\rm
NUV}\sim 0.25$. On the other hand, in the X-ray reprocessing model, the NUV
luminosity is proportional to the X-ray luminosity and scales as $L_{\rm NUV}
\propto L_{\rm X}^{0.5}$ (van Paradijs \& McClintock 1994), i.e., $\tau_{\rm
X}/\tau_{\rm NUV}\sim 0.5$.  The observed low ratio of $\tau_{\rm X}/\tau_{\rm
NUV}\sim 0.2-0.38$ is consistent with that the NUV emission is dominated by the
radiation of the outer non-irradiated viscous disc, and the viscous time-scale
is $\tau_{\rm \dot{M}} \sim 60$ days.

The emission at the given radius on the outer standard disc is characterized by
a (quasi) blackbody of temperature,
\begin{equation}
T(R)=\{\frac{3 G M_{\rm BH} \dot{M}}{8 \pi R^{3} \sigma}[1-(\frac{R_{\rm
\ast}}{R})^{1/2}]\}^{1/4},
\end{equation}
where the black hole mass $M_{\rm BH} = 10 M_{\odot}$ and the innermost stable
circular orbit (ISCO) of black hole $R_{\rm \ast} = 10^{7}$ cm are used in our
work. For an observer at a distance $D$, the flux at frequency $\nu$ from the
outer standard disc is
\begin{equation}
F_{\nu}=\frac{4 \pi h \cos i \nu^{3}}{c^{2} D^{2}}\int_{R_{\rm tr}}^{R_{\rm
disc}}\frac{R dR}{e^{h \nu / k T(R)}-1},
\end{equation}
where $i$ is the binary inclination (Frank et al. 2002). As discussed above
that the whole disc is thermally stable during the transient, we can put a
lower limit on the accretion rate by assuming $T(R_{\rm disc}) = 10^{4}$ K at
the onset of light curve transition. Using equations (2) and (3), we calculate
the values of the transition radius $R_{\rm tr}$ and accretion rate $\dot{M}$
for different values of $R_{\rm disc}$ under the condition of $L_{\rm NUV} =
10^{33}$ erg s$^{-1}$ and $T(R_{\rm disc}) = 10^{4}$ K. Figure 3 shows that the
outer standard disc is truncated at hundreds to thousands of $R_{\rm \ast}$ and
$\dot{M}\gg 10^{16}$ g s$^{-1}$ $\sim 10^{-3} \dot{M}_{\rm Edd}$ ($\dot{M}_{\rm
Edd} = 1.39\times10^{18}\times\frac{M}{M_{\odot}}$ g s$^{-1}$) at the onset of
light curve transition. Considering the viscous time-scale $\tau_{\rm \dot{M}}
\sim 60$ days, a lower limit for peak accretion rate $\dot{M}_{\rm peak} \sim
0.03 \dot{M}_{\rm Edd}$ is deduced from the exponential decay formula.

The observed $L_{\rm X}^{\rm peak} \sim 10^{35}$ erg s$^{-1}$ points to a very
low X-ray radiative efficiency $\eta=\frac{L_{\rm X}^{\rm peak}}{\dot{M}_{\rm
peak}c^{2}} < 5 \times 10^{-4}$, that is much lower than the value predicted
from ADAFs ($\eta > 0.01$ for $\dot{M}_{\rm peak} \sim 0.03 \dot{M}_{\rm Edd}$,
Xie \& Yuan 2012), indicating that a high proportion of accretion mass is lost
in outflows before reaching the central BH. Note that both the radio detection
during the outburst analyzed here (Sivakoff et al. 2011) and the quiescent
optical/infrared observations of this source (Shahbaz et al. 2013) suggested
the existence of a compact jet, supporting significant outflows from this
system. Such radiatively inefficient outflows were suggested to exist in other
BH LMXBs (e.g., Cyg X-1 and XTE J1118+480). Gallo et al. (2005) found that the
large-scale ring-like structure around Cyg X-1 was inflated by the inner jet,
and the kinetic energy of dark outflows dominated over radiations. Yuan et al.
(2005) fitted the most complete spectral energy distribution (SED; taken in
2000) of XTE J1118+480 with the hot accretion flow model including a jet, and
found that the power lost in the outflow exceeded the X-ray and radio emissions
by two orders of magnitude. The X-ray re-processing decreases as the X-ray
radiative efficiency decreases; on the other hand, the radiative efficiency of
the viscous energy release of the outer standard disc is relatively stable
depending on $R_{\rm tr}$ only. Eventually, the X-ray reprocessing is
overwhelmed by the outer standard disc in the NUV regime.

\begin{figure}
\includegraphics[width=9cm]{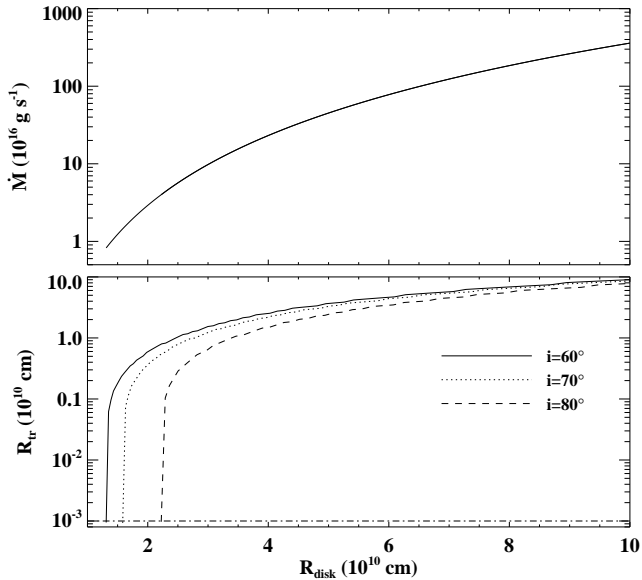}
\caption{Upper pannel: The accretion rate $\dot{M}$ vs. the disc size $R_{\rm
disc}$ at the onset of light curves transition. Bottom pannel: The transient
radii $R_{\rm tr}$ between the standard disc and ADAF for different values of
the disc size $R_{\rm disc}$. The horizontal dot-dashed line represents the
ISCO of black hole.\label{radii}}
\end{figure}

\section{Discussion and conclusion}

In this work, we perform a detailed analysis on both X-ray and NUV light curves
of Swift J1357.2-0933 during its 2011 outburst, and obtain the following
conclusions.

1. {\it Decay time-scales.} The light curves at all wavelengths display a
near-exponential decay in the early phase and transit to a faster-decay at late
times. The e-folding decay time of X-ray is $\tau_{\rm X}\sim 30$ days, while
the time-scales of longer-wavelength light curves are $\tau_{\rm NUV}\sim
80-150$ days, corresponding to a viscous time-scale of $\tau_{\rm \dot{M}} \sim
60$ days. This value is similar to that found in bright LMXBs (Chen et al.
1997), implying a similar viscosity parameter $\alpha$ in bright LMXBs and
VFXTs. The time scale in the late faster-decay stage is of a few days.

2. {\it NUV emission.} We firmly rule out the X-ray re-processing scenario and
confirm that the NUV emission is dominated by the viscous energy release of the
outer standard disc because of the high ratio of NUV luminosity to X-ray
luminosity ($\frac{L_{\rm NUV}}{L_{\rm X}} \sim 1$).

3. {\it Stable RIAFs.} When the irradiation is negligible, the DIM expects the
presence of re-flares and a sharp decline (on a thermal time-scale) in the last
evolution phase, which conflict with the observations analyzed here. Thus, the
outburst duration and recurrence times may not provide valid constraints on the
mean mass transfer rates in VFXTs. In contrast, the near-exponential decay
profile and the long decay time-scales indicate that and the accretion flow is
stable during the observations but with very low X-ray radiative efficiency.
Adopting the truncated disc model, we obtain a lower limit of peak accretion
rate $\sim 0.03 \dot{M}_{\rm Edd}$ and the X-ray radiative efficiency $\eta <
5\times10^{-4}$, which decreases as the luminosity declines. We stress that the
low efficiency is not just because of advection but also outflows, and our
model may also work for VFXTs which contain NSs.

As pointed out by Shahbaz et al. (2013), the distance of Swift J1357.2-0933 is
very uncertain, ranging between 0.5 and 6.3 kpc; therefore, its distance might
be significantly different from 1.5 kpc used above. We recalculate $R_{\rm tr}$
and $\dot{M}$ for different values of distance under the condition of $L_{\rm
NUV} = 1.0 \times10^{33}\times(\frac{D}{1.5 {\rm kpc}})^{2}$ erg s$^{-1}$ and
$T(R_{\rm disc}) = 10^{4}$ K. In our model, the obtained $\dot{M}$ is not
sensitive to the distance; however, the X-ray luminosity increases with $D$.
Therefore, the inferred X-ray radiative efficiency increases with $L_{\rm X}$
and also $D$. Note that the classification of bright, faint, and very faint
X-ray transients based on $L_{\rm X}^{\rm peak}$ is somewhat arbitrary. As a
matter of fact, the BH LMXB XTE J1118+480 has a brighter X-ray luminosity
($L_{\rm X} \sim 2\times10^{36}$ erg s$^{-1}$) but shares a lot of
observational properties with Swift J1357.2-0933, e.g., located at high
latitude, short orbital period, spectral type (Gelino et al. 2006; Shahbaz et
al. 2013). If Swift J1357.2-0933 is at 5 kpc, the inferred X-ray luminosity
$L_{\rm X} \sim 10^{36}$ erg s$^{-1}$, the accretion rate $\geq 0.03
\dot{M}_{\rm Edd}$, and the X-ray radiative efficiency $\eta < 5\times10^{-3}$
at the peak of outburst (Figure 4) are close to the SED fitting results of XTE
J1118+480 (Yuan et al. 2005). Swift J1357.2-0933 can be treated as a cousin of
XTE J1118+480.

At present, it is still unclear how the companion supplies such high accretion
rate. It was considered that the companion star can be heated by the accretion
disc, and the mass-transfer rate would be significantly enhanced when the
irradiation dominates the emission of the companion star, (e.g. Lasota 2001;
Viallet \& Hameury 2007). It is worth noting that the intrinsic emission of the
secondary is extremely faint. Thus, if the accretion disc emitted a moderate
X-ray luminosity before the outburst, its heating might be much greater than
the intrinsic emission of the companion star and trigger the outburst. However,
there is lack of observations before the outburst and a plausible mechanism for
the mass-transfer instability, and the origin of the high accretion rate is
still an open question.

\begin{figure}
\includegraphics[width=9cm]{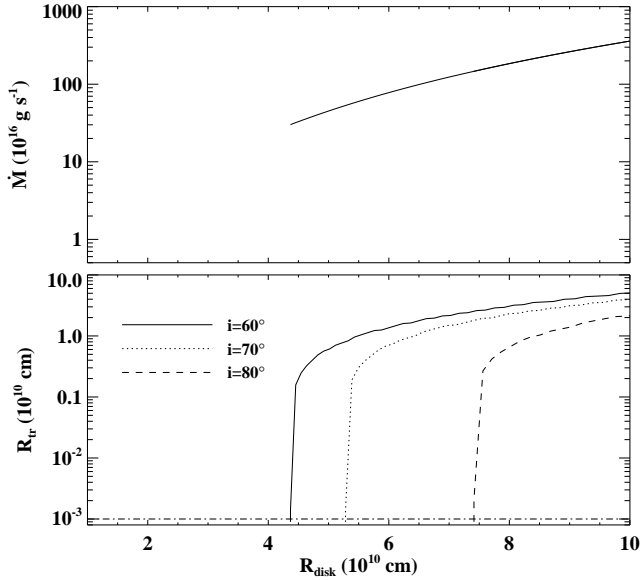}
\caption{The same as Figure 3 but with $d = 5 kpc$ under the conditions of
$L_{\rm NUV} = 1.1 \times10^{34}$ erg s$^{-1}$ and $T(R_{\rm disc}) = 10^{4}$
K. \label{d5kpc}}
\end{figure}

4. {\it Faster-decay stage.} The long decay time-scale in the late faster-decay
stage can not be explained by the irradiation effect nor the thermal
instability. We propose a possibility that the faster decay behavior can be
interpreted as due to the outer standard disc receding from the central BH.
Since the radial velocity in ADAFs is large, its accretion time-scale is
negligible. Therefore, the viscous time-scale is determine by the size of outer
standard disc $\tau_{\rm \dot{M}} \sim \frac{R_{\rm disc}-R_{\rm
tr}}{\bar{v}_{\rm r}}$, where $\bar{v}_{\rm r}$ is the mean radial velocity in
the outer standard disc. If the transition radius $R_{\rm tr}$ expands outward
at the velocity of $\bar{v}_{\rm r}$, the viscous time-scale becomes shorter,
leading to a quicker depletion of accretion disc mass and a faster-decay of
X-ray luminosity. In the meantime, less NUV emission is produced from the
smaller area (due to the increase of $R_{\rm tr}$) of the outer standard disc.
If $R_{\rm tr}$ increases by $\sim (0.1-0.2) R_{\rm disc}$ within $\sim 10$
days, the NUV emission decreases by a factor of a few. This model predicts that
the light curves of X-ray and NUV (quasi)-simultaneously transit from an
exponential decay to faster-decay, which can be checked by denser
multi-wavelength observations in the future.

\section*{Acknowledgements}

We would like to thank the referee for helpful suggestions and comments that
improved the clarity of the paper. S.S.W. thanks Weimin Gu for many valuable
suggestions. This work is partially supported with funding by 973 Program of
China under grant 2014CB845802, the National Natural Science Foundation of
China under grants 11133002, 11373036, and 11303022, the Qianren start-up grant
292012312D1117210, and by the Strategic Priority Research Program ``The
Emergence of Cosmological Structures'' of the Chinese Academy of Sciences,
Grant No. XDB09000000. S.S.W. is funded by the Co-Circulation Scheme, supported
by the EC-FP7 Marie Curie Actions-People-COFUND and T\"{U}B\.{I}TAK.


\end{document}